\documentclass{rspublic}

\pdfoutput=1

\usepackage{graphicx}
\usepackage{natbib}
 \usepackage{mathptmx}      

\begin{document}

\title[]{How does the Earth system generate and maintain thermodynamic disequilibrium and what does it imply for the future of the planet?}

\author[A. Kleidon]{Axel Kleidon}

\affiliation{Biospheric Theory and Modelling Group,\\ Max-Planck-Institut f\"ur Biogeochemie, \\ Hans-Kn\"oll-Str. 10, 07745 Jena, Germany}

\label{firstpage}

\maketitle

\begin{abstract}
{habitability, free energy, thermodynamics, global change, geoengineering}
%
The Earth's chemical composition far from chemical equilibrium is unique in our solar system and this uniqueness has been attributed to the presence of widespread life on the planet.
Here I show how this notion can be quantified by using non-equilibrium thermodynamics.
Generating and maintaining disequilibrium in a thermodynamic variable requires the extraction of power from another thermodynamic gradient, and the second law of thermodynamics imposes fundamental limits on how much power can be extracted. 
%
%
With this approach and associated limits, I show that the ability of abiotic processes to generate geochemical free energy that can be used to transform the surface-atmosphere environment is strongly limited to less than 1 TW.  Photosynthetic life generates more than 200 TW by performing photochemistry thereby substantiating the notion that a geochemical composition far from equilibrium can be a sign for strong biotic activity.
Present-day free energy consumption by human activity in form of industrial activity and human appropriated net primary productivity is in the order of 50 TW and therefore constitutes a considerable term in the free energy budget of the planet.
When aiming to predict the future of the planet, we first note that since global changes are closely related to this consumption of free energy, and the demands for free energy by human activity are anticipated to increase substantially in the future, the central question in the context of predicting future global change is then how human free energy demands can increase sustainably without negatively impacting the ability of the Earth system to generate free energy. 
This question could be evaluated with climate models, and the potential deficiencies in these models to adequately represent the thermodynamics of the Earth system is discussed.  I then illustrate the implications of this thermodynamic perspective by discussing the forms of renewable energy and planetary engineering that would enhance overall free energy generation and thereby "empower" the future of the planet.
\end{abstract}

\section{Thermodynamic disequilibrium as a sign of a habitable planet}

In the search for easily recognizable signs of planetary habitability, Lovelock (\citeyear{Lovelock1965}) suggested the use of the chemical disequilibrium associated with the composition of a planetary atmosphere as a sign for presence of widespread life on a planet.  He argued that Earth's high concentration of oxygen in combination with other gases, particularly methane, constitutes substantial chemical disequilibrium that would quickly be dissipated by chemical reactions if it were not continuously replenished by some processes.  Since life dominates these exchange fluxes, he argued that life is the primary driver that generates and maintains this state of chemical disequilibrium in the Earth's atmosphere (see also \citet{Lovelock1975}).  

Atmospheric composition is only one aspect of the Earth system that is maintained far from a state of equilibrium.  Another example of disequilibrium is the atmospheric water vapor content, which is mostly far from being saturated.  If it were not for the continuous work being performed in form of dehumidification by the atmospheric circulation \citep{PauluisHeld2002a,PauluisHeld2002b}, the atmosphere would gain moisture until it is completely saturated, as this is the state of thermodynamic equilibrium between the liquid and gaseous state of water.  Topographic gradients on land also reflect disequilibrium as erosion would act to deplete these gradients into a uniform equilibrium state, and life plays an important role in the processes that shape topographic gradients \citep{DietrichPerron2006,DykeEtal2010}.

The maintenance of these disequilibrium states seems to contradict the fundamental trend towards states of thermodynamic equilibrium, as formulated by the second law of thermodynamics.  To understand why this is fully consistent and even to be expected from the laws of thermodynamics, we need to resort to the formulations of non-equilibrium thermodynamics and formulate Earth system processes and their interactions on this basis.  While Lovelock's further research, e.g. on the Gaia hypothesis \citep{LovelockMargulis1974} and the Daisyworld model \citep{WatsonLovelock1983} has contributed substantially to the emergence of Earth system science that considers the functioning of the Earth as one, interconnected system \citep{Lovelock2003b,SchneiderBoston1991,SchneiderEtal2004}, the use of thermodynamics as a basis for the holistic integration of processes of the Earth system is practically absent from mainstream Earth system science.  If we had such a basis, Lovelock's conjecture could easily be evaluated and we could evaluate how human activity alters the fundamental nature of the planet.  Such a basis would seem critical to have to guide in the process of managing the impacts of human activity within the Earth system in the future.

 This paper lays out how non-equilibrium thermodynamics can be used to develop a holistic view of how disequilibrium is generated and maintained within the Earth system, what this view would imply for the effects of human activities on the Earth system, and what potential deficits there are in the numerical models that we use to assess Earth system change.  To do so, the paper is structured along a series of questions.  First, a brief overview of non-equilibrium thermodynamics is provided to address the question of how disequilibrium is generated and maintained without violating the second law of thermodynamics.  In essence, it is shown how free energy is generated from one thermodynamic gradient and transferred to another, causing disequilibrium in thermodynamic variables that are not directly related to heat and entropy.  More specifically, it is shown how work is derived (or power generated) from the planet's external forcing and initial conditions that then fuels a hierarchy and cascade of free energy generation, transfer, and dissipation.  One can imagine this to be similar to an engine that is fueled by heating and that can drive a series of belts and wheels, resulting in motion and cycling of mass.  Then, the fundamental constraints that limit the extent of disequilibrium are addressed, which leads to the Carnot limit, its limited applicability to Earth system processes, and its extension to the maximum power principle and the proposed principle of Maximum Entropy Production (For a reader familiar with the thermodynamic basis, sections \ref{sec:generation} - \ref{sec:limits} can easily be skipped).  This is followed by an evaluation of the generation rates of free energy of the present-day Earth system to estimate the relative importance of the drivers of present-day disequilibrium states, particularly regarding chemical disequilibrium.  The impacts of human activities are then discussed in this context and their likely consequences for planetary disequilibrium and free energy generation.  After explicitly discussing some potential deficits in current Earth system models regarding the dynamics of free energy, the paper closes with a brief summary and conclusions.

\section{How is disequilibrium generated and maintained?}
\label{sec:generation}

Typically, textbook teaching of thermodynamics mostly deals with isolated systems that are maintained in a state of thermodynamic equilibrium, characterized by a maximum in entropy.  This is the consequence of the second law, since processes can only increase entropy in an isolated system.  To understand the generation and maintenance of disequilibrium -- and the associated low entropy within the system --, we note that the Earth system is not an isolated system, so that the nature of its thermodynamic state is substantially different.  A state of disequilibrium does not violate the laws of thermodynamics, but is rather a consequence of these.  In the following, a brief overview of the thermodynamics away from equilibrium is given to provide the basics of generating and maintaining disequilibrium and its relation to free energy generation within a system.  The following overview is not meant to be exact and encompassing, but rather illustrative and explanatory.  For more details, the reader is referred to textbooks on non-equilibrium thermodynamics \citep[e.g.][]{KondepudiPrigogine1998,LebonEtal2008}.

\subsection{Defining the Earth as a thermodynamic system}
Even though it seems somewhat formal, we first need to define the boundaries of the Earth system.  The way we choose the boundaries is, in theory, arbitrary.  Depending on how we choose the boundary, we may get a different type of thermodynamic system in that we need to consider different types of exchange fluxes through the boundaries.  By choosing the boundaries well, we can make our description of the thermodynamic system a lot easier because we may need to consider fewer exchange fluxes to describe the system.

Three types of thermodynamic systems exist: (i) isolated systems are systems in which no exchange of energy and mass takes place with the surroundings; (ii) closed systems exchange energy, but no mass with the surroundings; and (iii) open systems that exchange both, energy and mass.  

When we deal with the Earth system, a good choice for the boundary is the top of the atmosphere.  There, the dominant exchange is radiative, with low entropy solar radiation -- in terms of its photon composition as well as its confinement to a narrow solid angle -- entering the Earth system, and terrestrial radiation with some scattered solar radiation being returned to space.  With this choice of boundary, the Earth is almost a closed system (ignoring the relatively small exchange due to gravity and mass, such as hydrogen escape to space that could have played an important role in the Earth's past).

Note that subsystems of the Earth are typically placed at some thermodynamic meaningful boundaries as well.  For instance, the separation of the climate system into the atmosphere, ocean and land follows the boundaries defined by the different states -- gaseous, liquid, and solid.  When we deal with the boundary of the biosphere, we would place it at the interface of organic, living matter to its inorganic, non-living surroundings.  The shape of this boundary would be rather complex.  Because the exchange of mass between these subsystems is substantial, these subsystems are examples of open thermodynamic systems.

\subsection{The first and second law of thermodynamics}
Once the boundary is defined, we need rules to determine the limits on how exchange fluxes at the boundary can be altered.  This leads us to the first and second laws of thermodynamics.  These laws express the constraints on the rate at which work can be extracted from a heat gradient to generate free energy, and it provides the direction of natural processes towards states of higher entropy.  These two laws form the basis to understand what is needed to drive and maintain a state of thermodynamic disequilibrium.  

The first law balances the change in internal energy $dU$ within a system with the amount of heat exchange $dQ$ with the surroundings and the amount of work done by the system $dW$:
\begin{equation}
\label{eqn:firstlaw}
dU = dQ - dW
\end{equation}
In principle, the internal energy $U$ of a system represents mostly the amount of stored heat, but other types of energy also contribute to this term (see below; these contributions are generally small).  When we deal with the Earth system, however, work is not extracted from the system, but is performed within the system itself and, hence, $dW$ is not taken away from $U$.  This work is, for instance, reflected in the acceleration or lifting of mass, that is, it is taken away from the heat content to generate a non-heat form of free energy.  So when we want to use the first law to quantify generation rates of free energy, we first need to distinguish between heat vs. non-heat contributions to $U$ (with $U_{heat}$ referring to the contribution of heat to $U$).  For completeness, we also need to consider that the free energy generated by the system is also eventually dissipated, so that an additional, dissipative heating term $D$ needs to be considered in the balance of $U_{heat}$.  So when we consider changes through time, we need to balance the changes in heat $dU_{heat}/dt$ with the rates of net heating of the system due to external forcings, $J_h = dQ/dt$, the extracted power $P = dW/dt$ to forms of free energy, and the dissipative heating $D$ resulting from irreversible processes within the system that dissipate free energy:
\begin{equation}
\label{eqn:firstlawintime}
\frac{dU_{heat}}{dt} = J_h - P + D
\end{equation}
While eqn. \ref{eqn:firstlawintime} looks like a typical climatological energy balance, it is more than that because it includes the transfer and depletion rates of free energy related to $P$ and $D$ associated with other forms of free energy.  The magnitudes of $P$ and $D$ are typically quite small in comparison to $J_h$ and therefore not important for the heat balance that determines the changes in temperature.  However, $P$ and $D$ are critical for the dynamics of the planet as these are driven by free energy generation and dissipation and are intimately linked to the maintenance of the disequilibrium state, as we will see further below.  

The second law states that the entropy $S$ of an isolated system can only increase, i.e, $dS \geq 0$.  The change in entropy $dS$ is defined as the amount of heat $dQ$ added or removed at the temperature $T$ of the system:
\begin{equation}
\label{eqn:entropy}
dS = \frac{dQ}{T}
\end{equation}
To demonstrate how the diffusion of heat is a manifestation of the second law, let us consider two reservoirs of heat with different temperatures $T_h$ and $T_c$ with $T_h > T_c$ within a system.  The difference in temperatures would drive a diffusive heat flux that would remove some amount of heat $dQ$ from the warmer reservoir and add it to the colder reservoir.  The entropy of the warmer reservoir would decrease by $dQ/T_h$, while the entropy of the colder reservoir would, once $dQ$ is fully mixed, increase by $dQ/T_c$.  The entropy of the whole system would change by $dS = -dQ/T_h + dQ/T_c$, and $dS > 0$ because $T_h > T_c$.  When this heat exchange is sustained through time, resulting in a diffusive heat flux $J = dQ/dt$, then the diffusive heat exchange produces entropy $\sigma$ at a rate of 
\begin{equation}
\label{eqn:entropyproduction}
\sigma = \frac{dS}{dt} = J \cdot \left(\frac{1}{T_c} - \frac{1}{T_h}\right)
\end{equation}
When we deal with a non-isolated system, then we need to also consider the entropy changes due to the exchange fluxes of energy and mass with the surroundings.  Then, eqn. \ref{eqn:entropy} becomes a balance equation for the entropy of the system which relates the changes in entropy $dS/dt$ with the entropy production $\sigma$ by irreversible processes within the system, e.g. heat diffusion, and the net exchange of entropy across the system boundary, $J_{s,net}$.  The net entropy exchange captures the fact that the energy that is exchanged with the surroundings is not of the same quality.  For instance, it captures the fact that solar, shortwave radiation is of different composition than terrestrial, longwave radiation that leaves the Earth.  Hence, the entropy balance is written as:
\begin{equation}
\label{eqn:entropybalance}
\frac{dS}{dt} = \sigma - J_{s,net}
\end{equation}
with an isolated system represented by $J_{s,net} = 0$.

Note that when we deal with the system and its surroundings together, e.g. Earth and space, then we still deal with an isolated system.  The total entropy of the system plus its surroundings then increases by $J_{s,net}$, which is consistent with the second law of thermodynamics \citep{OzawaEtal2003,LineweaverEgan2008}.

\subsection{Quantifying thermodynamic disequilbirium}
Thermodynamic disequilibrium and the balance equation for free energy $A$ are directly linked to the first and second law (eqns. \ref{eqn:firstlawintime} and \ref{eqn:entropybalance}).  Free energy is generated when the power extracted from heating ($P$ in eqn. \ref{eqn:firstlawintime}) is used to perform the work of building up some other gradient, which then stores free energy in some other form.  The change of free energy $dA$ is given by the balance of added power $P$, which adds free energy, and the rate of dissipation $D$ by some irreversible process, which depletes free energy:
\begin{equation}
\label{eqn:freeenergy}
\frac{dA}{dt} = P - D
\end{equation}
Dissipation is linked to both: to the energy balance through the dissipated heat, and to the entropy balance by the entropy that this dissipation produced ($D$ is thus linked to $\sigma$ in eqn.  \ref{eqn:entropybalance}).  Not considered in eqn. \ref{eqn:freeenergy} is the potential transfer of free energy from one type to generate free energy of another type.  This aspect is dealt with further below when discussing free energy generation and transfer within the Earth system.

The free energy $A$ in a system is directly related to the distance to thermodynamic equilibrium.  To relate the two, we consider the first law (eqn.  \ref{eqn:firstlaw}) and note that the added heat $dQ$ either contributes to the internal energy of the system $dU$ or to the free energy generated $dA$ by performing the work $dW$.  We can then express the change of entropy $dS$ (as in eqn. \ref{eqn:entropy}) as the sum of two contributions, $dS = dS_{heat} + dS_{diseq}$, relating to the change in internal energy $dU/T$ and to the change in free energy $dA/T$ within the system.  Hence, the disequilibrium of a system, as expressed by $S_{diseq}$, is directly related to the free energy content $A$ by $dS_{diseq} = dA/T$.

In steady state, $P = D$.  When the rate of dissipation is expressed as $D = A/\tau$, then the steady-state free energy within a system is $A_{ss} = P \cdot \tau$ and the associated disequilibrium $S_{diseq,ss} = P \cdot \tau/T$.  The extent of disequilibrium (and mean free energy) within a system is hence directly related to how much power $P$ is contained in the processes that generate free energy.  Since the disequilibrium is also related to the timescale of its depletion $\tau$, the extent of observed disequilibrium does not necessarily tell us how active the system is in generating and dissipating free energy.

\subsection{Illustration with a simple model}
\begin{figure}[htbp]
\noindent
\centerline{\includegraphics[width=25pc]{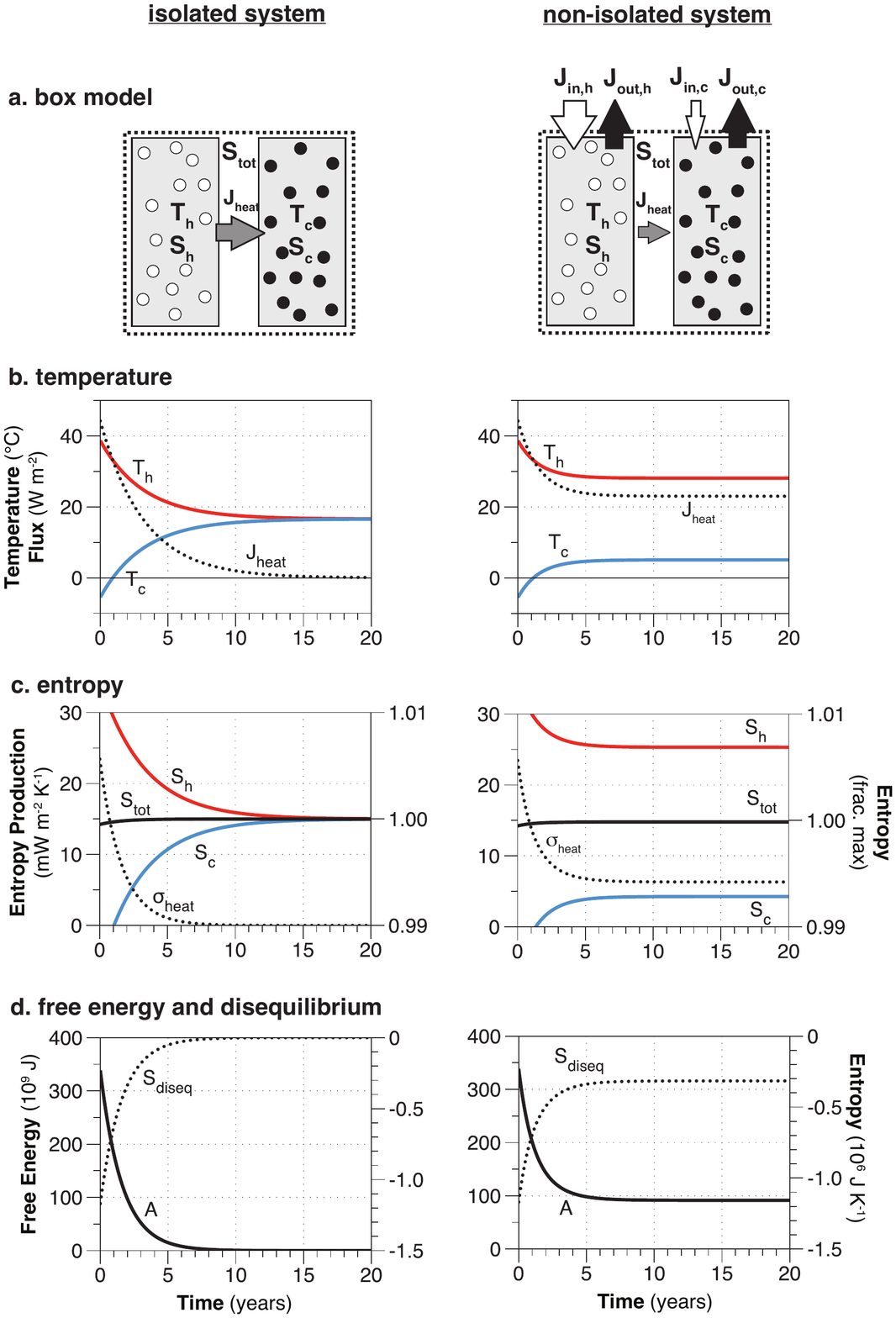}}
\caption{Simple box models to demonstrate the time evolution of an isolated (left column) and non-isolated (right column) thermodynamic system.  a: layout of the box model with $T_h$, $T_c$: temperatures of the hot and cold reservoir respectively; $J_{heat}$ heat flux; $S_h$, $S_c$, $S_{tot}$: entropy of the hot, cold reservoir and the whole system respectively;  $J_{in,h}$, $J_{in,c}$, $J_{out,h}$, $J_{out,c}$: heat exchange fluxes with the surroundings (subscripts refer to in: heating from the surroundings, out: cooling to the surroundings, h: hot reservoir, c: cold reservoir).  b: time evolution of temperatures and heat flux $J_{heat}$ between the boxes.  c: entropies $S_h$, $S_c$, and $S_{tot}$ and entropy production $\sigma$ by heat exchange.  d:   free energy $A$ and disequilibrium entropy $S_{diseq}$.  Assembled from \citet{Kleidon2010a}.}
\label{fig:thermodiseq}
\end{figure}

The concepts of free energy and disequilibrium for two types of thermodynamic systems are demonstrated using the two simple systems shown in Fig. \ref{fig:thermodiseq}.  The two systems consider two heat reservoirs that are linked by a heat flux and have initially the same, uneven distribution of heat.  The only difference between the system is that the system on the right column of Fig. \ref{fig:thermodiseq} exchanges heat with its surroundings.  For the details of the model, see Kleidon (2010).

The system on the left of Fig. \ref{fig:thermodiseq} is an isolated system (Fig. \ref{fig:thermodiseq} left column).  In this system, the difference in heat content is dissipated through time, as shown by the equilibriation of the temperatures (Fig. \ref{fig:thermodiseq}b left).  The heat flux decreases with the declining temperature difference, as is the associated entropy production (Fig. \ref{fig:thermodiseq}c left).  While the entropy of the colder box increases, the entropy of the warmer box decreases due to the removal of heat.  The entropy of the whole system increases due to the equilibriation, which is clearly seen in the depletion of $A$ and $S_{diseq}$ (Fig. \ref{fig:thermodiseq}d left).

The system on the right of Fig. \ref{fig:thermodiseq} is a closed system, in which both heat reservoirs exchange heat with the surroundings.  The initial difference in heat content also declines in this system, but it reaches a steady-state away from equilibrium with a non-vanishing difference due to the differential heating that the two reservoirs receive (Fig. \ref{fig:thermodiseq}b right).  We observe a steady-state heat flux $J_{heat}$ that aims to deplete the temperature difference, but cannot deplete it to zero due to external, differential heating of the system.  Hence, the fluxes through the boundaries play a critical role for the maintenance of disequilibrium within the system, as reflected by a difference in the entropies between the reservoirs (Fig. \ref{fig:thermodiseq}c right), non-zero free energy $A$ and disequilibrium $S_{diseq}$ (Fig. \ref{fig:thermodiseq}d right).  Feedbacks of the dynamics of the system onto the exchange fluxes at the boundaries and constraints that shape the flexibility of these determine how far a system can evolve and maintain itself away from a state of thermodynamic equilibrium.

\section{How is disequilibrium and free energy generated within the Earth system?}
\label{sec:Earth}

We now take the simple formulations of the previous section and apply them to the Earth system.   As discussed above, the exchange of the Earth system with its surroundings is accomplished primarily by the exchange fluxes of solar and terrestrial radiation.  These radiative fluxes result in radiative heating and cooling when absorbed or emitted.  The associated radiative heating and cooling fluxes, $J_{heat}$ and $J_{cool}$, and the energy balances that reflect the temperatures at which this heating and cooling takes place, $T_{heat}$ and $T_{cool}$, form the constraints for physical forms of free energy generation by heat engines as formulated by the first and second law of thermodynamics.

Note that these heat engine constraints do not apply directly to photochemical means of free energy generation.  Solar radiation contains a significant amount of free energy that is embedded in the photon composition, which reflects the high emission temperature of the Sun.  When solar radiation enters the Earth system, its flux density is much diluted compared to when it was emitted, so that its photon composition is far from equilibrium with its respective radiative equilibrium temperature and equilibrium photon composition.  A simple way to illustrate the free energy in solar radiation is to consider the temperature gradient between the emission temperatures of the Sun and the Earth, which would yield a Carnot efficiency of $\Delta T/T \approx (5760K - 255K)/5760K \approx$ 95 \% (which is only an approximation, as detailed analyses would need to consider the entropy of radiation; for the explanation of the Carnot limit, see below).  This disequilibrium, however, is not provided in form of a temperature gradient, but rather in form of the associated photon composition.  Hence, it can only be exploited by processes that involve photons transformations, such as absorption, but not by heat engines.

Photochemistry is one way that utilizes the free energy of solar radiation.  It utilizes excited electrons that have absorbed solar photons in the visible range.  Electronic absorption can generate electronic free energy and thereby avoid that the energy of solar photons is directly converted into heat after absorption.  Photochemistry can in principle yield substantially more free energy than heat engines driven by the same radiative fluxes, but it requires suitable photochemical mechanisms to exploit this free energy.  By performing photosynthesis, life does exactly this and generates substantial amounts of chemical free energy at the Earth's surface.  Note that the photochemistry taking place in the Earth's stratosphere also generates free energy by converting molecular oxygen into ozone, but this free energy is rapidly dissipated, and is therefore unlikely to provide substantial rates of free energy that is available to drive geochemical reactions at the surface.

We deal with this contribution later and focus first on the constraints imposed by physical transfer processes related to heat engines and what these imply for the functioning of the planet.

\subsection{Planetary balance equations}
The dynamics of the planet are constrained by the associated global balance equations for total internal energy $U_{planet}$, heat $U_{planet,heat}$, free energy $A_{planet}$ (with $U_{planet} = U_{planet,heat} + A_{planet}$), and entropy $S_{planet} = S_{heat} + S_{diseq}$:
\begin{equation}
\frac{dU_{planet}}{dt} = J_{heat} - J_{cool}
\end{equation}
\begin{equation}
\frac{dU_{planet,heat}}{dt} = J_{heat} - J_{cool} - P_{total} + D_{total}
\end{equation}
\begin{equation}
\frac{dA_{planet}}{dt} = P_{total} - D_{total}
\end{equation}
\begin{equation}
\frac{dS_{planet}}{dt} = \frac{dS_{heat}}{dt} + \frac{dS_{diseq}}{dt} = \frac{J_{heat}}{T_{heat}} - \frac{J_{cool}}{T_{cool}} + \sigma_{total}
\end{equation}
where $P_{total}$ is the total power (or generation rate of free energy) of the Earth system, $D_{total}$ is the total dissipation of free energy, and $\sigma_{total}$ the total entropy production by irreversible processes.  Note that the change in disequilibrium, $dS_{diseq}/dt$,  is directly related to the change in free energy, $dA_{planet}/dt$, by $dS_{diseq}/dt = -1/T_{planet} \cdot dA_{planet}/dt$ (as above, with a representative temperature $T_{planet}$), and the total dissipation $D_{total}$ contributes to the total entropy production $\sigma_{total}$ (although entropy is also produced by diffusive processes that are not related with the dissipation of free energy, so that $\sigma_{total} \ge D_{total}/T$).

Disequilibrium and free energy of the planet can result from the depletion of planetary initial conditions and by exploiting the fluxes at the planetary boundary.  Both of these have, of course, to be consistent with the second law, which requires $\sigma_{total} \ge 0$ in the global balance equations above.  To understand how disequilibrium and free energy in other forms of energy than heat is generated, we need to include these in our considerations of the total internal energy $U_{planet}$.  Other forms of energy, such as gravitational energy, kinetic energy, binding energy, chemical energy, and so on, can be expressed in terms of conjugate variables, pairs of thermodynamic variables that taken together express different forms of energy.  Changes in the total internal energy $dU_{planet}$ can then be expressed as the sum of changes in heat and changes in the energy of the other forms:
\begin{equation}
\label{eqn:duplanet}
dU_{planet} = d(T S) + \sum_i d(p_i v_i) + \sum_i d(\phi_i M_i) + \sum_i d(\mu_i M_i) + \sum_i d(A_i \xi_i) + \cdots
\end{equation}
Here, the first term on the right side expresses the contribution by the heat content of the system (temperature $T$ and entropy $S$, yielding $U_{planet,heat}$), by kinetic energy (momentum $p_i$ and velocity $v_i$), by gravitational energy (gravitational potential $\phi_i$ and mass $M_i$), by binding energies (chemical potential $\mu_i$ and mass $M_i$) and chemical energy (affinities of the reactions $A_i$ and extent of the chemical reactions $\xi_i$).  The sums in eqn. \ref{eqn:duplanet} run over the different contributions to a particular form of free energy, e.g. the contributions of air flow, water flow (oceans and rivers), mass flow of plate tectonics and mantle convection, and so on to the total kinetic energy within the system.  Other forms of energy, e.g. electric or magnetic energy, can be expressed in a similar fashion (see e.g. \citet{KondepudiPrigogine1998,Alberty2001}), but are neglected here for simplicity.

\subsection{Dynamics of free energy and disequilibrium}
The change in total internal energy $dU_{planet}$ does not tell us much about the dynamics that take place within the Earth system and the associated disequilibrium. When one form of energy is converted into another, the total internal energy $dU_{planet}$ does not necessarily change since this is just a conversion between different contributions to $dU_{planet}$.  When, for instance, free energy in some chemical compound is depleted by an exothermic chemical reaction, then the free energy associated with chemical compounds is reduced (i.e. $d(A_i \xi_i) < 0$), which is balanced by the increase in the heat content (i.e. $d(T S) > 0$).  Likewise, when a gradient in heat results in the generation of kinetic energy, then the term $d(T S)$ decreases and the term $d(p_i v_i)$ increases, but $dU_{planet}$ is unaffected.  When atmospheric motion lifts dust, the kinetic energy fuels the generation of potential energy of the dust grains, i.e. $d(p_i v_i) < 0$ to result in $d(\phi_j m_j) > 0$.  Ultimately, the kinetic energy is dissipated by friction, that is, $d(p_i v_i) < 0$ and $d(T S) > 0$, which again is not associated with a change $dU_{planet}$.  Hence, the energy balance does not inform us about how much free energy is generated, dissipated, and transferred between conjugate variables and how much disequilibrium is being maintained.

What is not captured by the energy balance is that work is required to generate free energy and drive associated dynamics.  For instance, work is needed to accelerate mass to convert a gradient in heat into motion.  Or, more generally, the dynamics of conversions are driven by the generation and depletion of gradients of the different sets of conjugate variables, where the depletion of one gradient fuels the generation of another gradient.   These dynamics are described by the generation and depletion rates of the different forms of free energy that all contribute to $A_{planet}$.  

The conversions among different forms of energy cannot take place arbitrarily at any rate or direction.  The second law requires that $\sigma_{total} \ge 0$.  It thereby constrains the entropy balance, and imposes limits on the direction and the rates of conversions.  In steady state, the entropy balance yields $J_{s,net} = \sigma_{total}$, i.e. the total rate of entropy production by all irreversible processes within the Earth system is constrained by the entropy exchange at the system boundary.  For the Earth system, these are the exchange fluxes of solar and terrestrial radiation.  Hence, the entropy balance is intimately linked with the energy balance, specifically to the net entropy exchange associated with $J_{heat}$ and $J_{cool}$.  

Furthermore, the entropy balance is linked to the free energy balance.  First, the rates of free energy generation, $P$, are restricted by the entropy balance, specifically by the requirement of the second law of $J_{s,net} = \sigma_{total} \ge 0$.  This aspect will be dealt with in more depth in the next section in connection to how the entropy balance limits power generation to a characteristic maximum possible rate.  Second, the dissipated free energy $D$ is connected with the total entropy production within the system $\sigma_{total}$.  Hence, the overall dynamics of free energy generation and the resulting disequilibrium state are closely connected with the net entropy exchange at the system boundary.

\subsection{Planetary hierarchy of free energy generation, transfer, and dissipation}

\begin{figure}[htbp]
\noindent
\centerline{\includegraphics[width=25pc]{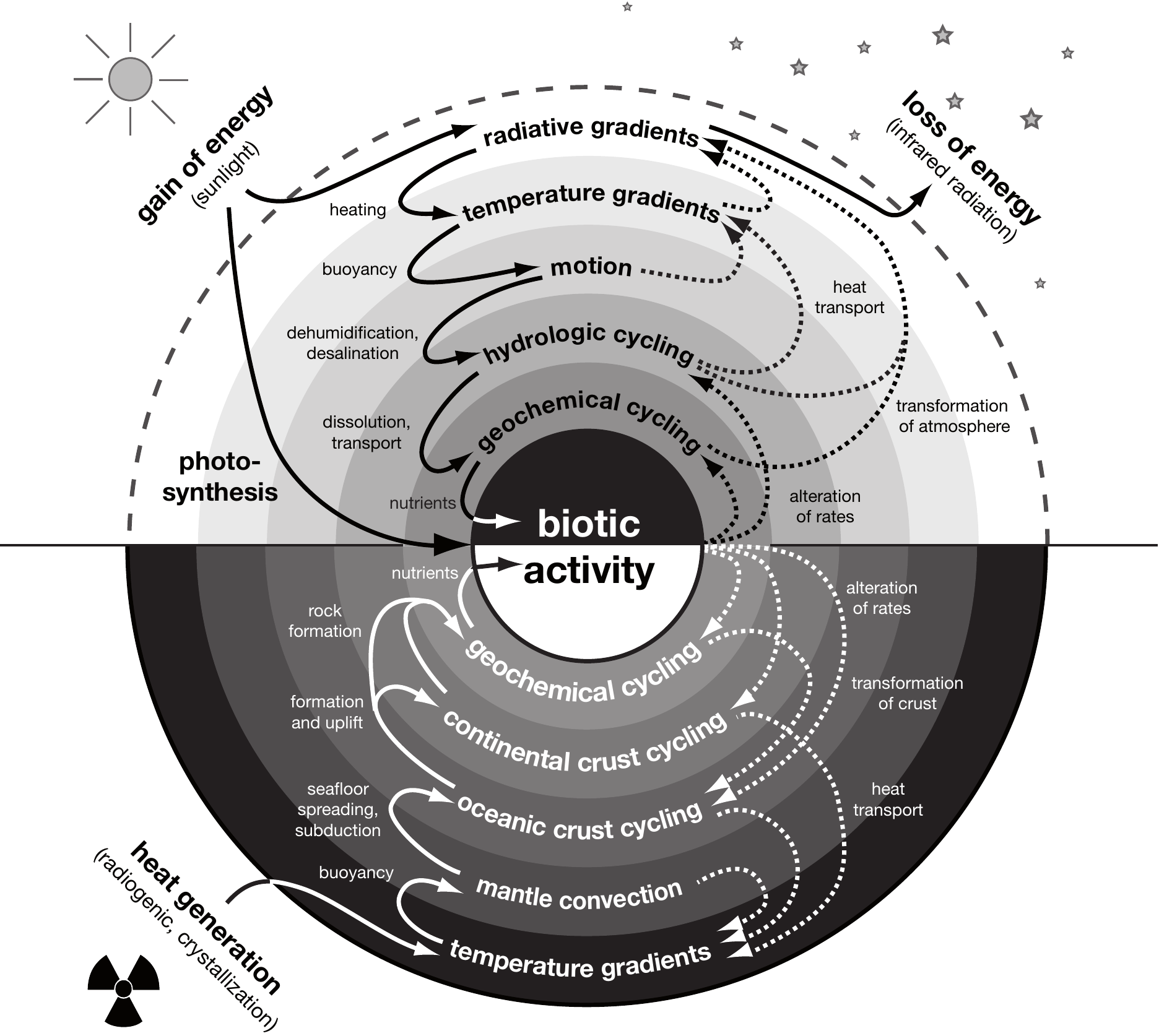}}
\caption{Schematic diagram of the planetary hierarchy of free energy generation, transfer and dissipation (solid lines) and associated effects (dotted lines).  The different layers are associated with different forms of free energy and gradients associated with disequilibrium.  For instance, motion is associated with gradients in momentum and represent kinetic energy. Hydrologic cycling is associated with gradients in chemical potential and geopotential and is associated with potential and chemical free energy.  From \citet{Kleidon2010a}.}
\label{fig:wholeEarth}
\end{figure}

When we consider most of the abiotic processes within the Earth system, these are ultimately driven by the free energy generated by heat engines driven by external forcings or initial conditions.  The power extracted from the differential heating is then converted further into some other forms.  Hence, understanding the maintenance of disequilibrium within the Earth system needs to be viewed in an encompassing, holistic perspective of free energy generation and transfer among different processes, as shown in simplified form in Figure \ref{fig:wholeEarth}.  

This holistic view, which was developed in \citet{Kleidon2008c,Kleidon2009c,DykeEtal2010,Kleidon2010a,Kleidon2010b}, shows two major drivers for heat engines that fuel abiotic free energy generation.  The first driver results from the spatial and temporal variation in solar radiation at the system boundary.  This flux generates gradients in temperature, which causes changes in air density, pressure gradients, and results in motion.  Atmospheric motion powers the dehumidification of the atmosphere, which generates the free energy that drives evaporation and desalination of seawater at the surface and condensation aloft, and hence the global cycling of water.  The transport of water on land provides the free energy in form of potential free energy to drive river runoff and sediment transport, and in form of chemical free energy associated with freshwater to chemically dissolve the continental crust.  Hence, we get a sequence of transformations from a radiative energy flux that results in a small fraction of free energy generation that can alter the geochemical nature of the surface.

The second driver is associated with the depletion of the initial conditions of the planet at formation, in form of secular cooling of the interior, heating by radioactive decay, and crystallization of the core.  The differential heating results in a similar sequence of free energy generation that results in plate tectonics, uplift of continental crust, and generation of geochemical free energy at the surface \citep{DykeEtal2010}.

The distribution of free energy generated by the heat engines across different types of free energy results in a different way to look at interactions and feedbacks within the Earth system (dotted lines in Fig. \ref{fig:wholeEarth}).  The extent to which power is transferred from one form of energy to another obviously affects the dynamics associated with both forms of energy.  For instance, atmospheric motion is driven by pressure and temperature gradients, but this motion results in the transport of mass and heat that depletes these gradients.  When power is removed from motion, e.g. to lift moisture, then it strengthens water cycling at the expense of slowing down motion.  Large-scale hydrologic cycling also depletes the driving force of motion through the transport of latent heat.  Hence, this holistic view places interactions and feedbacks into the context of how these affect the rates of power generation as the primary driver for Earth system dynamics.  Given that there are characteristic maximum rates of power generation for many processes, as we will see in the following, these place upper bounds on to the strength of interactions within the Earth system.  
 
\section{What are the limits to the generation of disequilibrium and free energy?}
\label{sec:limits}

The maximum of free energy generation in classical thermodynamics is characterized by the Carnot limit.  In the following, I briefly explain how the Carnot limit is derived in order to understand the implicit assumptions being made.  Then, the Carnot limit is applied to a slightly different setting more representative of natural processes to show that this yields the maximum power principle and/or Maximum Entropy Production (MEP) for complex, non-equilibrium systems such as the Earth system.

\subsection{The Carnot limit}

The Carnot limit is derived directly from the first and second law.  It considers a system as shown in Fig. \ref{fig:maximumpower}a, where a heat flux $J_{in}$ adds heat to the system while the heat flux $J_{out}$ removes heat from the system.  The first law in steady state then tells us that   
\begin{equation}
\label{eqn:energy_carnot}
0 = J_{in} - J_{out} - J_{ex}
\end{equation}
where $J_{ex} = P_{ex}$ is the extracted power from the system.  The constraint on the maximum value of $P_{ex}$, and consequently the Carnot limit, originates from the entropy balance of the system.  The entropy exchange of the system is set by a fixed temperature gradient, with the influx of heat taking place at a corresponding temperature of $T_{in}$, and the outflux at a temperature of $T_{out}$.  Hence, the entropy balance of the system in steady state is given by:
\begin{equation}
\label{eqn:entropy_carnot}
0 = \sigma + \frac{J_{in}}{T_{in}} - \frac{J_{out}}{T_{out}}
\end{equation}
Because $P_{ex}$ is free energy, it is not associated with an entropy, i.e. it can be completely used for performing work, and therefore does not show up as a term in eqn. \ref{eqn:entropy_carnot}.  The best case for extracting work is when no irreversible process takes place within the system and the entropy production within the system is zero (i.e. $\sigma = 0$).  This condition, $\sigma = 0$, is the limit of what is permitted by the second law.  In this case, the entropy balance can be used to express $J_{out}$ as a function of $J_{in}$, $T_{in}$ and $T_{out}$, and the first law (eqn. \ref{eqn:energy_carnot}) yields an expression for the maximum rate $P_{ex,max}$ at which work can be extracted from the system:
\begin{equation}
P_{ex,max} = \frac{(T_{in} - T_{out})}{T_{in}} \cdot J_{in}
\end{equation}
This is the well-known Carnot limit, with an associated thermodynamic efficiency defined as $\eta = (T_{in}-T_{out})/T_{in}$.  

Note that the derivation of the Carnot limit contains important assumptions.  First, it assumes that the temperature gradient is fixed, i.e. there is no effect of the rate of extracted work on the temperatures at the system boundary.  Second, it assumes that no other irreversible processes take place.  If $\sigma > 0$ because of some unavoidable irreversible process taking place within the system, the extractable work would need to be less than the Carnot limit.  And third, the balance of free energy is not in a steady state since the free energy generated by the extracted work is not dissipated, and the related waste heat is not added back to the system.  These assumptions cannot be made for Earth system processes.  Interactions play a critical role, and heat fluxes often deplete the gradients by which these are driven.  Also, processes compete for the same driving gradient, e.g. convection and conduction, hence the assumption of no other irreversible processes cannot be made.  Finally, the Earth system is closed in that extracted work is dissipated in steady state.

We next consider what the maximum power limit is when these assumptions are relaxed.

\subsection{The maximum power limit}

\begin{figure}[htbp]
\noindent
\centerline{\includegraphics[width=25pc]{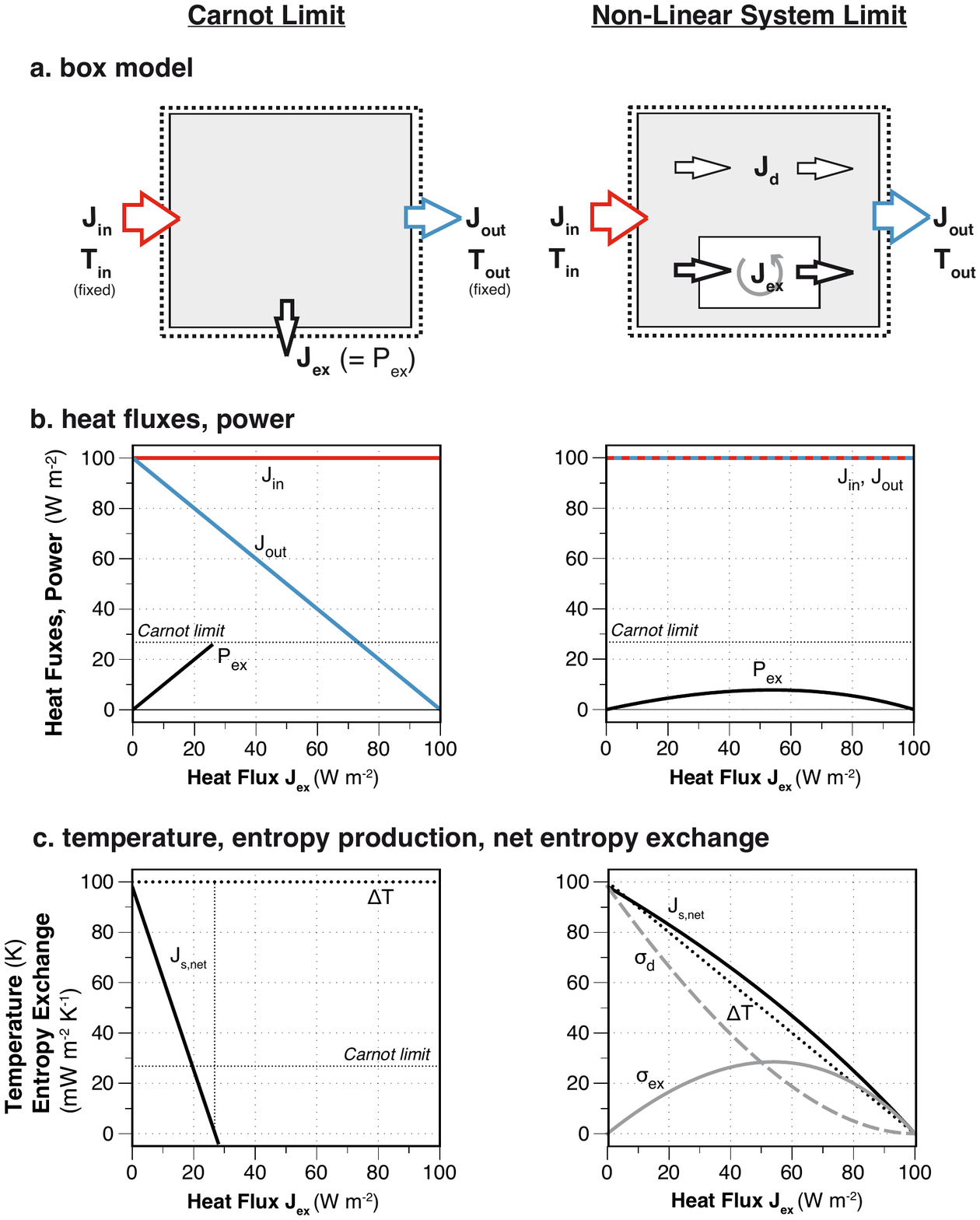}}
\caption{Illustration of the maximum extractable power in the common Carnot limit (left column) and in a non-linear system (right column) in which power extraction competes with another dissipative process (e.g. diffusion, radiative exchange) and the temperature gradient is affected by the extracted power and the associated dynamics.  a: definition of system boundary and fluxes.  In addition to an influx of heat $J_{in}$, an outflux $J_{out}$, and an extracted heat flux $J_{ex}$, the right system also has a diffusive heat flux $J_{d}$ that competes with $J_{ex}$ for the depletion of the temperature gradient.  b: heat fluxes and power vs. extracted heat flux $J_{ex}$.  The Carnot limit is shown by the dashed line for comparison.  c: temperature difference $\Delta T$ and net entropy exchange $J_{s,net}$ vs. extracted heat flux $J_{ex}$.  The Carnot limit corresponds to the extracted heat flux $J_{ex}$ at which $J_{s,net} = 0$ (left column).  In case of the system shown in the right column, the maximum in extracted power corresponds closely to a maximum in entropy production $\sigma_{ex}$ associated with power extraction.  See \citet{Kleidon2010a} for model details.}
\label{fig:maximumpower}
\end{figure}

The assumptions of the Carnot limit can be relaxed in a slightly more complex setting as shown in Fig. \ref{fig:maximumpower}a (right).  In this setting, free energy generation competes with diffusive dissipation of the temperature gradient.  The extracted work is dissipated and the resulting heat is added back into the system.  When expressing the diffusive process as $J_d = k_d (T_{in} - T_{out})$, and applying the Carnot limit to the extracted heat flux $J_{ex}$, we obtain for the work extraction $P_{ex}$:
\begin{equation}
P_{ex} = \frac{(T_{in} - T_{out})}{T_{in}} \cdot J_{ex} = \frac{(J_{in} - J_{ex}) \cdot J_{ex}}{k_d T_{in}}
\end{equation}
where the expression for $J_d$ and the energy balance constraint $J_{in} = J_d + J_{ex}$ were used to substitute $(T_{in} - T_{out})$ and $J_d$.  It is easy to show that this expression has a maximum for a heat flux $J_{ex,opt} = J_{in}/2$ of (if the dependency of the $T_{in}$ in the denominator is neglected, see also \citet{Kleidon2010a})
\begin{equation}
\label{eqn:maxpower}
P_{ex,max} = \frac{1}{4} \cdot \frac{(T_{in,0} - T_{out,0})}{T_{in,0}} \cdot J_{in}
\end{equation}
where $T_{in,0}$ and $T_{out,0}$ are the temperatures at the system boundary in the absence of work extraction (i.e. $J_{ex} = 0$).  The reduction to 1/4 compared to the Carnot efficiency if it were applied directly to $J_{in}$ can be understood as follows:  at the state of maximum power, the temperature gradient is depleted to half its value, because the extracted work enhances the overall heat flux.  The competition with diffusive loss yields another reduction by a factor of 2.  This results in the maximum power limit of eqn. \ref{eqn:maxpower}, which is equivalent to the well known maximum power principle in electrical engineering.

The state of maximum power extraction coincides very closely to the state at which the entropy production is at a maximum (Fig. \ref{fig:maximumpower}).  This maximum in entropy production relates to the proposed principle of Maximum Entropy Production (MEP, \citet{OzawaEtal2003,KleidonLorenz2005,MartyushevSeleznev2006,KleidonEtal2010}), which states that complex systems with sufficient degrees of freedom maintain a steady state at which entropy production is maximized.  There are some theoretical developments to support the general nature of this principle and relate it to the Maximum Entropy formalism of equilibrium thermodynamics \citep{Dewar2003,Dewar2005a,Dewar2005b,Niven2009}.  While the derivation of the maximum power limit above merely establishes an upper bound -- just as the Carnot limit --, the proposed MEP principle would imply that complex systems would actually evolve to and maintain such a steady state -- just like an engineer would work towards achieving the Carnot limit when designing an engine.

There are several indications that natural processes, such as turbulence \citep{OzawaEtal2001}, convection \citep{OzawaOhmura1997}, and the atmospheric circulation on Earth \citep{Paltridge1975,Paltridge1978,KleidonEtal2003,KleidonEtal2006} and other planetary systems \citep{LorenzEtal2001,Lorenz2010}, operate close to states of maximum entropy production and/or maximum power/dissipation (see also reviews by \citet{OzawaEtal2003,KleidonLorenz2005,MartyushevSeleznev2006,Kleidon2009a}).  This would suggest that the dynamics of complex systems are indeed characterized by maximization of power and dissipation.

\subsection{Maximum power of material processes}

The above derivation of the maximum power limit is based on the extraction of work from a heating gradient.  Equivalent derivations can be made for material fluxes and associated forms of energy.  In order to find states of maximum power, we need to identify a trade-off between the force that drives the flux, and the flux that depletes the force.  In the above example of maximum power derived from a heating gradient, the force is associated with the temperature gradient that drives the convective heat flux.  This flux turn depletes the force by depleting the temperature gradient, resulting in the limit of extractable power.

The maximum power limit is well established within electrical engineering.  Here, the trade-off exists between the resistance associated with a load that is connected to a generator with an inherent, internal resistance.  The greater the resistance of the load, the greater the gradient in electric potential across the load, but because of the higher overall resistance, the current is reduced.  Since the electric power drawn by the load is the product of current and potential gradient across the load, a maximum exists for the power that can be drawn from the electric generator.

To demonstrate an equivalent tradeoff that is more relevant for the Earth system, take the transfer of power contained in some flow (e.g. air flow, river flow) to perform work on a material flux (e.g. dust transport, sediment transport).  The momentum gradient between the flow and the surface exerts a drag force that performs the work of lifting and accelerating particulates to the speed of the flow.  The greater the removal of power from the flow, the more work can be performed on the particulates, but at the cost of a reduced flow velocity and thereby a reduced ability to transport particulates.  Hence, a trade-off exists between the force, which is proportional to the momentum gradient, and the resulting flux of particulates, which depletes the force.  Consequently, maximum power limits should be a rather general feature of Earth system processes.  These limits constrain the extent to which power can be transferred within the Earth system as shown in Fig. \ref{fig:wholeEarth}, and to the extent to which thermodynamic variables can be maintained at states of thermodynamic disequilibrium.  

\subsection{Maximum power and feedbacks}

These upper bounds and the associated state of maximum power are relevant for how the Earth system responds to perturbations.  If the emergent dynamics are organized in such a way that they maximize power generation, then these dynamics would evolve towards the maximum power state after any perturbation.  Hence, maximization is inherently associated with the system reacting to perturbations with negative feedbacks \citep{OzawaEtal2003}.  

If the nature of the boundary conditions change, this would alter the conditions to which the maximization is subjected to.  In this case the emergent dynamics would evolve in such a way as to maximize power under the altered boundary conditions.

\subsection{Maximum power and disequilibrium}

When we want to relate these limits in power generation to the original motivation of understanding the drivers for disequilibrium within the Earth system, we note first that the state of maximum power depends on boundary conditions only, and not on the material properties of the process under consideration.  This can be seen by eqn. \ref{eqn:maxpower}, which depends on the temperatures and the heat flux, both of which merely describe the conditions at the boundary.  It does not depend on e.g. the density or viscosity of the fluid.  The associated disequilibrium, however, does depend on material properties.  The disequilibrium and the associated free energy content results from the balance of generation and dissipation of free energy.  While the maximum state in generation is described by the maximum power state of the boundary conditions, the dissipation is intimately linked with material properties such as density and viscosity of the fluid in the case of motion.  Hence, a state of maximum power is not necessarily equivalent with states of maximum disequilibrium or maximum free energy content.  As it would seem that the maximization of power (or entropy production) is based on a more general and better justified basis, I focus on power and free energy generation rather than disequilibrium and free energy content in the following.

\section{What are the generation rates of free energy for the present-day Earth?}
\label{sec:presentday}

\begin{figure}[htbp]
\noindent
\centerline{\includegraphics[width=25pc]{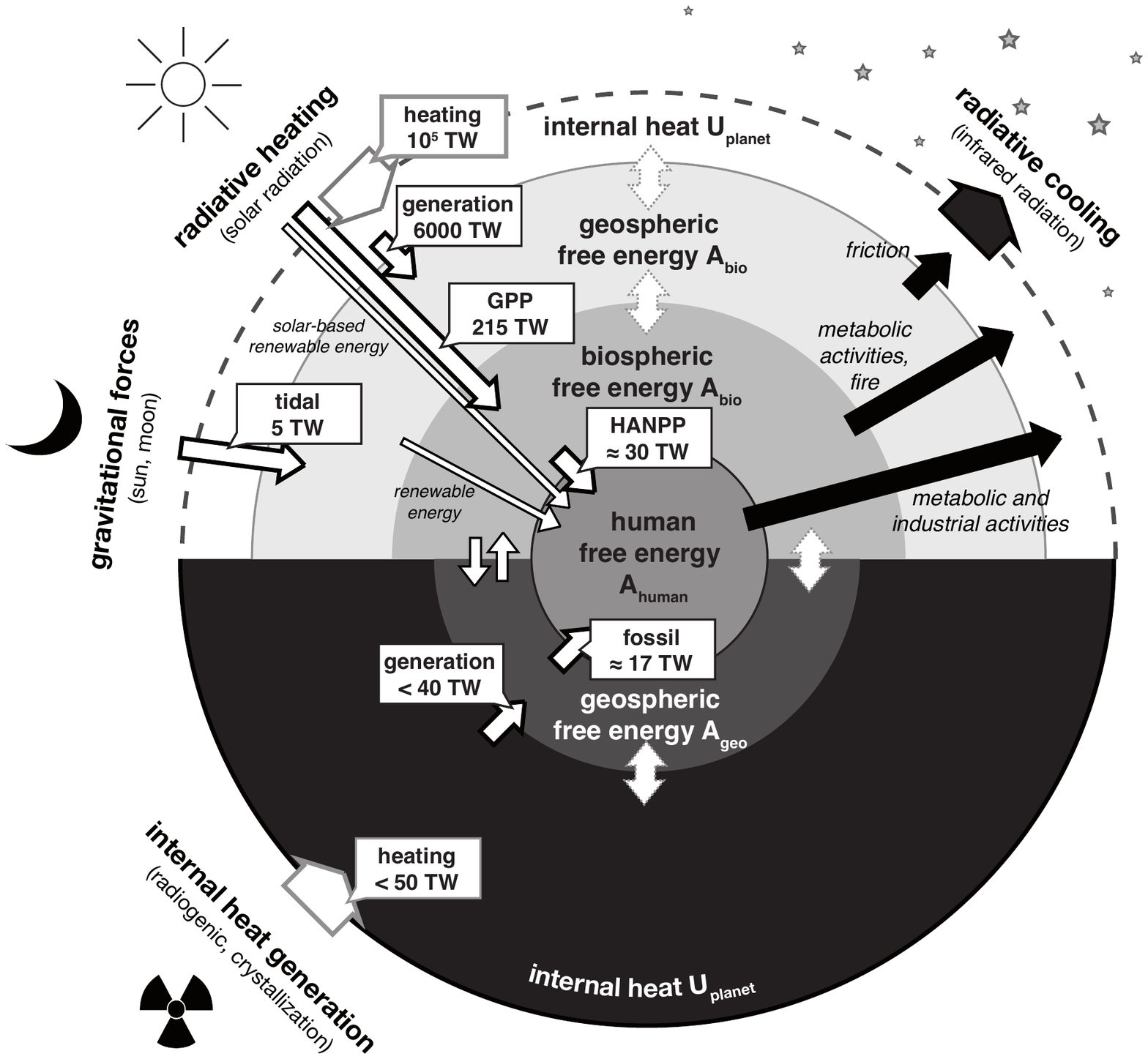}}
\caption{Estimates for global free energy generation rates at the planetary scale.  Heating rates by solar radiation and interior processes are shown in boxes with grey border.  Free energy generation processes are shown by white boxes and arrows with black borders ("tidal": free energy generation by tidal forces; "generation": free energy generation by heat engine processes; "GPP": gross primary productivity, i.e. chemical free energy generation by photosynthesis).  Human activity is driven by about 30 TW of human appropriation of net primary productivity of the biosphere ("HANPP") and the consumption of fossil fuels ("fossil"). 1 TW = 10$^{12}$ W.  Estimates taken from \citet{Kleidon2010a}.}
\label{fig:freeenergy}
\end{figure}

To understand the drivers for present-day disequilibrium, we need to estimate the generation rates of free energy within the Earth system.  These can be estimated by considering the primary drivers that supply free energy from external sources and that feed the hierarchy of transfer shown in Fig. \ref{fig:wholeEarth}.  Using these drivers, a global free energy budget is derived and shown in Fig. \ref{fig:freeenergy}.  In contrast to the well-established global energy balance, the free energy balance emphasizes the importance of the biota in the planetary free energy generation (in particular in form of chemical free energy) and highlights the magnitude of human activity in dissipating free energy.  The estimates are based on \citet{Kleidon2010a} and are described in the following.

\subsection{Drivers of free energy generation}
Four major drivers are responsible for free energy generation within the Earth system:  

\begin{itemize}
\item{\textit{solar heat engines:} incident solar radiation is associated with spatial and temporal variations that maintain temperature gradients and fuel heat engines.  These gradients are the main driver for climate system processes.  The total free energy generated from this source includes the generation of potential free energy associated with air, water vapor, and aerosols, the kinetic energy associated with motion in the atmosphere, oceans, and river flow, the chemical free energy generated by dehumidification and desalination of sea water, the electric free energy generated by thunderstorms and so on.  All of these are fueled by uneven heating and cooling, resulting in vertical and horizontal gradients in density and pressure.  We can estimate the maximum power available to maintain these types of free energy by considering the radiative forcing at the boundary using simple considerations \citep{Kleidon2010a}.  Absorption of solar radiation at the surface and cooling by emission of radiation aloft generates a vertical gradient in heating that can be converted into free energy.  Using a mean surface heating of 170 W m$^{-2}$ and typical temperatures of 288K and 255K, \citet{Kleidon2010a} estimates that free energy generation from this vertical gradient is less than 5000 TW (note that much of this power is used for vertical mixing and transport and is likely to contribute relatively little to large-scale cycling and transport).  Due to the planet's geometry and rotation, absorption of incident solar radiation result in horizontal gradients.  Using a mean difference in solar radiation of about 40 \% between the tropics and the poles yields an upper limit of about 900 TW.  The temporal variation of heating in time yield an additional power of about 170 TW at maximum efficiency, so that the total power generated from radiative heating gradients is in the order of 6000 TW;}

\item{\textit{solar photochemical engines}: incident solar radiation contains wavelengths that can be used to generate chemical free energy when visible or ultraviolet radiation is absorbed by electronic absorption or photodissociation.  Photodissociation can, in principle, generate radicals that are associated with free energy, but it is omitted here since those compounds have very short residence times and therefore unlikely to result in sustained free energy generation of significant magnitude.  Photosynthesis is able to generate longer-lasting free energy by using complex photochemistry that prevents rapid dissipation.  Using typical values for global gross primary productivity and typical free energy content of carbohydrates yields a generation rate of chemical free energy of about 215 TW \citep{DykeEtal2010};}

\item{\textit{gravitational engines}:  gravitational forces by the Moon and the Sun provide some additional free energy by generating potential free energy mostly in the ocean in form of tides.  Estimates place the total generation rate at around 5 TW \citep{FerrariWunsch2009};}

\item{\textit{interior heat engines}: radioactive decay, crystallization of the core, and secular cooling of the interior provide means to generate free energy within the interior.  This free energy is associated with the kinetic energy of mantle convection and plate tectonics, with potential free energy generation associated with plate tectonics, with magnetic free energy generation associated with the maintenance of the Earth's magnetic field, and with geochemical free energy generation associated with metamorphosis and other geochemical transformations.  Given that the heat flux from the interior is less than 0.1 W m$^{-2}$ at the Earth's surface, maximum efficiency estimates by \citet{DykeEtal2010} yield a maximum generation rate of free energy in the various forms of about 40 TW.}

\end{itemize}

In summary, the total free energy generation for current conditions yield about $P_{geo,a} \approx 6070$ TW of power by physical processes within the atmosphere from the exchange fluxes at the Earth-space boundary, about $P_{bio} \approx 215$ TW of chemical free energy by photosynthesis, and about $P_{geo,b} \approx 40$ TW driven by the depletion of initial conditions in the interior.  Hence, the total power generation by the planet is about $P_{planet} = P_{geo,a} + P_{geo,b} + P_{bio} \approx 6325$ TW. 

\subsection{Free energy transformations and planetary geochemical disequilibrium}

Only a small fraction of the total generated power $P_{planet}$ is available for geochemical transformations.  Of the $\approx$ 6070 TW of geophysical free energy generated within the Earth's atmosphere, most is likely to be dissipated by atmospheric convection.  More certain are large-scale dissipative terms.  About 1000 TW are  dissipated by frictional dissipation by the large scale atmospheric circulation \citep{LiEtal2007}, 65 TW are transferred into the oceans to generate waves and maintain the wind-driven circulation \citep{FerrariWunsch2009}, about 560 TW are associated with lifting water to the height at which it condenses and precipitates to the ground \citep{PauluisEtal2000,Kleidon2010a}, and about 27 TW is associated with desalinating seawater.  

Geochemical free energy is generated by abiotic means mostly by hydrologic cycling.  Of the 560 TW involved in hydrologic cycling, about 13 TW are involved in maintaining streamflow.  Of these 13 TW, only some fraction can be used to mechanically transform the continental crust by transporting sediments to the ocean.  Precipitation on land also yields chemical free energy associated with disequilibrium of freshwater and the Earth's crust.  This chemical free energy can be used to dissolve minerals of the continental crust.  The power generated by precipitation is about 0.15 TW.  The contribution by interior processes to the generation of geochemical free energy is likely to be very small since most of the 40 TW of power is involved in the generation of kinetic energy associated with mantle convection and plate tectonics.

In contrast to these very small generation terms of chemical free energy, biotic productivity generates 215 TW of chemical free energy.  Not all of this free energy is available for geochemical transformations of the environment, as the metabolic activity of organisms consumes about half of the generated free energy.  This contribution is nevertheless likely to be 1 - 2 orders of magnitude larger than abiotic means of geochemical free energy generation.  Hence, this estimate substantiates the suggestion by Lovelock (\citeyear{Lovelock1965}) that the Earth's planetary geochemical disequilibrium is mostly attributable to the presence of widespread life on the planet.

\subsection{Free energy appropriation and dissipation by human activity}

It is well recognized that human activity substantially alters the planet, as reflected by the suggestion to refer to the current geologic era as the "anthropocene" \citep{Crutzen2002}.  The additional heating by the burning of fossil fuels is, however, minute with its approximately 17 TW of primary energy consumption \citep{EIA2009} compared to the solar heating in the order of 10$^{5}$ TW.   The planetary impact of human activity is much easier noticeable when considering the free energy appropriation related to human activity.  Free energy is needed for maintaining human activity as it fuels the basic metabolic requirements (mostly food production) as well as the industrial activities (mostly fossil fuel consumption).

The basic requirement for metabolic activity is met by food production and the associated human appropriation of net primary production (HANPP, \citet{VitousekEtal1986,RojstaczerEtal2001,ImhoffEtal2004,Kleidon2006a}) from the biosphere.  The concept of HANPP is more encompassing than merely the metabolic needs of humans, as it also considers contributions such as grazing by domesticated animals and the use of firewood.  It is used here as a first order approximation for the human demands for meeting metabolic energy needs.  This need for free energy is estimated to be about 10 - 55 \% of the net primary productivity on land \citep{RojstaczerEtal2001}.  Using the estimated 40 \% as a best guess \citep{VitousekEtal1986} and converting the annual net primary productivity into units of free energy, this yields a free energy appropriation of $P_{human,meta} \approx 30$ TW.  Note that this number is quite uncertain, not all of it is associated with food production, and 30 TW are not needed to directly meet human metabolic needs.  With a human population of about 5 billion, and a basal metabolic rate of 100W, this merely yields 0.5 TW of dissipation.  The difference between 0.5 TW and 30 TW could be explained by a relatively large inefficiency of converting biotic productivity into food.  If we, for instance, assume that the conversion efficiency is about 10 \% for converting the productivity of grass to the maintenance of livestock and for the conversion of livestock into meat production, then the energetic needs to meet human metabolic requirements would translate into an equivalent chemical free energy generation by biotic productivity in the order of 50 TW, so that the estimate of 30 TW seems reasonable, but would require a more careful analysis.  In addition, the industrial use of free energy, mostly from fossil fuels, is associated with $P_{human, ind} \approx 17$ TW of free energy.  

In total, human activity therefore consumes about $P_{human} = P_{human,meta} + P_{human, ind} \approx 47$ TW.  When compared to the planetary budget of free energy generation, especially regarding abiotic means of geochemical free energy generation of less than 40 TW,  human energy consumption is a substantial term in the budget.  The 47 TW of human consumption exceeds all free energy generated and consumed by geologic processes of less than 40 TW in the Earth's interior.

\section{How does human activity change planetary free energy generation?}
\label{sec:humanactivity}

The free energy used for human activities are, of course, drawn out of the Earth system and thereby affect its state.  At present, much of the free energy needs for industrial use are met by depleting a stock of geological free energy (in form of fossil fuels) and this results in global climatic change due to higher concentrations of carbon dioxide in the Earth's atmosphere.  If this depletion is going to be replaced by renewable sources of free energy -- as commonly suggested to avoid emissions of carbon dioxide --, then this is going to leave an impact on the free energy balance of the planet.  Other impacts of human activity, such as the emission of methane, nitrous oxide, or soot also relate, directly or indirectly, to the combustion of fuels or to food production, and should also relate to the Earth's free energy balance.  Hence, it would seem appropriate to relate human activity as well as its impacts on the Earth system to its basic driver, the need for free energy.  This need for free energy would seem to be the most important metric to measure the impact of humans on the planet and would seem to serve to be a highly useful metric to evaluate potential future impacts.  

\begin{figure}[htbp]
\noindent
\centerline{\includegraphics[width=36pc]{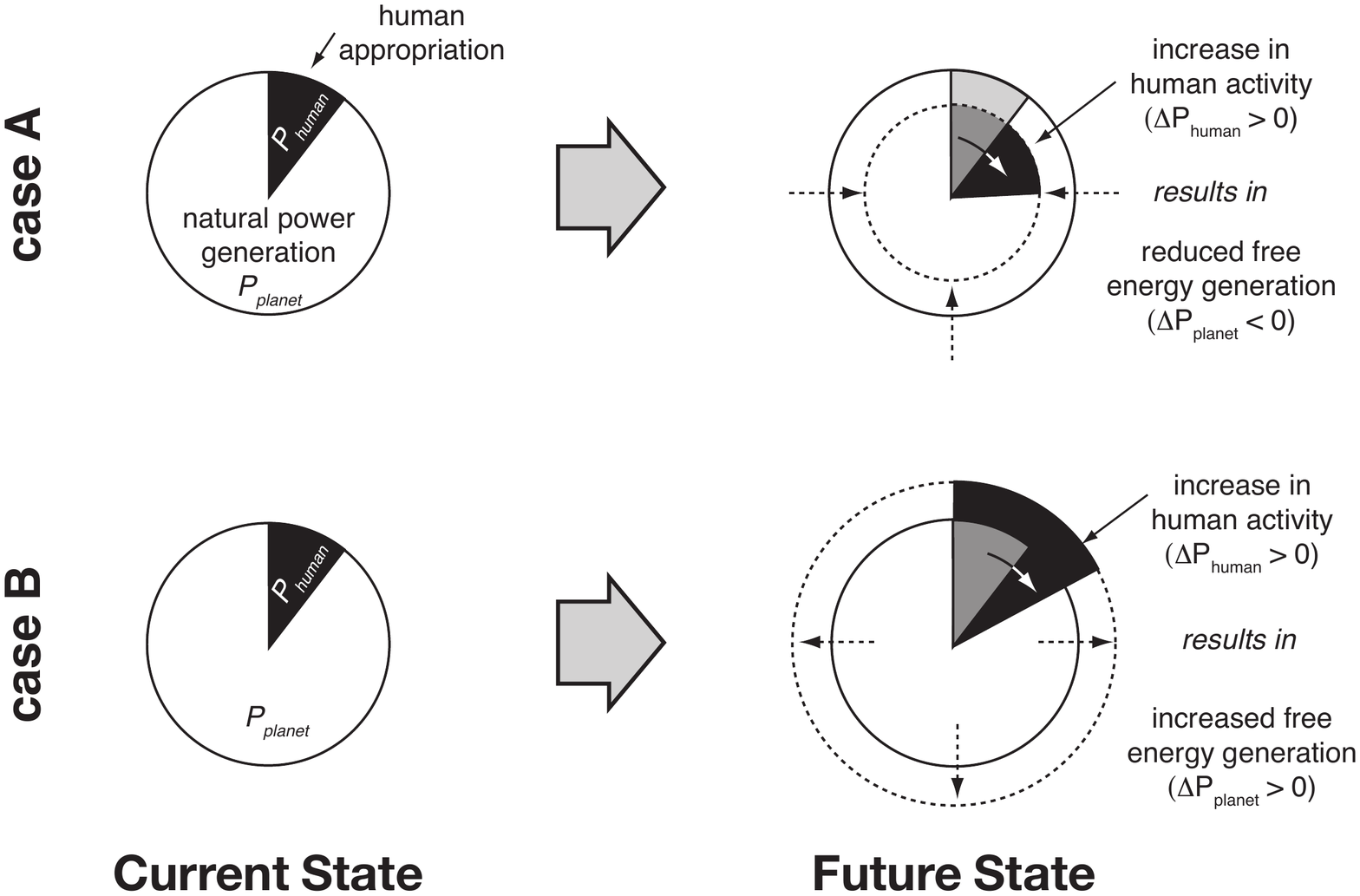}}
\caption{Schematic diagram of possible contrasting effects of an increase in human activity, corresponding to an increase in free energy consumption $\Delta P_{human} > 0$, on the free energy generation capability of the biosphere and the Earth system.  In case A (top), the effects of $\Delta P_{human} > 0$ result in a reduction of free energy generation by natural processes, i.e. $\Delta P_{planet} < 0$.  Case B (bottom) shows the alternative case in which the effects of $\Delta P_{human} > 0$ result in an overall enhancement of free energy generation by natural processes, i.e. $\Delta P_{planet} > 0$.  For a sustainable future with inevitable increases in energy demands by human activity it is argued in the text that human effects should follow case B.}
\label{fig:empoweringEarth}
\end{figure}

As we have already seen in the last section, human activity already consumes a considerable share of the free energy in relation to how much is generated within the Earth system.  When we think about the future state of the planet, it would seem almost inevitable that human activity will increase further, in terms of population size and standard of living, among others.  Both of these will require more free energy to sustain.  Then, the central question is going to be whether this increase in human activity is going to be met by degrading the ability of the Earth system to generate free energy or whether these demands will be met by enhancing the ability of the Earth system to generate free energy.  If we think of the free energy budget shown in Fig. \ref{fig:freeenergy} as a pie, then these questions amount to the issue whether an increase in human activity in the future is going to decrease or increase the planetary pie of free energy generation, thereby depleting or enhancing the planetary disequilibrium.  This concept is illustrated in Fig. \ref{fig:empoweringEarth}, and its application is illustrated qualitatively in the following.

The need for meeting our metabolic demands results in the conversion of natural lands into agricultural use.  Because humans appropriate productivity (i.e. $\Delta P_{human,meta} > 0$), the associated changes in land cover are likely to result in less productivity available for the biota (i.e. $\Delta P_{bio} < 0$), therefore resulting in less free energy generation to drive biotic activity.  The associated land cover changes result in different functioning of the land surface with consequences for the atmosphere.  A shift from forests to agricultural lands is usually accompanied with a higher reflectivity and a reduced ability to recycle water back into the atmosphere \citep{Bonan2008}.  When we consider the extreme scenario of all vegetation being removed from land, the associated climatic conditions would likely be less favorable to biospheric free energy generation, as demonstrated by extreme climate model simulations \citep{KleidonEtal2000,Kleidon2002}.  It would hence seem that an increase in agricultural activity ($\Delta P_{human,meta} > 0$) in principle would result in shrinking the pie of biotic free energy generation ($\Delta P_{bio} < 0$).  This decrease was also shown in sensitivity simulations with a coupled vegetation-climate model to the magnitude of human appropriation by \citet{Kleidon2006a}.

The need for heating and industrial activity associated with human activity, $P_{human,ind}$, is currently fueled to a large extent by fossil fuels.  Even though the fossil free energy is not taken away from an active, geological process that affects the Earth system at present, its consumption results in carbon dioxide emissions into the atmosphere.  The expected increase in energy demands $\Delta P_{human,ind} > 0$ as well as the necessary shift towards sustainable sources of free energy, such as wind power, hydropower, tidal power and so on, the appropriation of such sources of free energy by humans would obviously reduce the free energy within the system, i.e. $\Delta A_{planet} < 0$.  Since these processes are likely to operate already at maximum efficiency, the associated impacts are likely going to be such that the power of these processes are going to decrease as a result of human appropriation (i.e. $\Delta P_{planet} < 0$.  That this is indeed the case has been demonstrated with sensitivity simulations with a climate model to the magnitude of wind power extraction at the surface \citep{MillerEtal2011}.

Both examples of meeting the human demands for free energy suggest that human activity will result in detrimental effects in terms of the ability of the Earth system to generate free energy.  We can, however, also imagine another scenario.  If human activity is directed to have impacts of the sort that these would act to enhance free energy generation within the Earth system, as shown in Fig. \ref{fig:empoweringEarth}b, then this could have beneficial effects on the overall system in that the ability to generate free energy within the Earth system is enhanced.  Two examples are given below to illustrate such potentially positive impacts due to human activity.

To meet an increase in demands for metabolic free energy generation, one could imagine that if currently unproductive lands are utilized and converted into agricultural lands with the use of technology (e.g. by irrigation with desalinated seawater), then this could result in enhanced free energy generation by the biota.  Then, an increase $\Delta P_{human,meta} > 0$ would be met with a change $\Delta P_{bio} > 0$.  This could, for instance, be accomplished by using technology to "green" the desert (see also \citet{OrnsteinEtal2009}).  If the resulting gain in productivity is larger than the technological needs for free energy to enable desert greening, we would have a win-win situation of overall enhanced free energy generation by the Earth system.  If "desert greening" would indeed yield overall such an energetic gain, however, would need to be carefully evaluated in greater detail.

The increase in demands for industrial free energy could be met by more efficient use of solar radiation.  Currently, most of the solar radiation is absorbed and immediately converted into heat.  Differential heating is, however, is very inefficient in converting solar radiation into free energy.  Photovoltaic cells or the use of direct solar radiation are two examples of technologies that could be used to enhance the efficiency of converting solar radiation into free energy.  Their use in currently unvegetated regions, such as deserts, could result in an overall enhanced ability of the Earth system to generate free energy with the use of human technology -- just as in the example of "desert greening".  

Such conscious planning of human effects to enhance the power generation by the Earth system would constitute a form of "geoengineering" or "Earth system engineering".  Currently proposed measures of "geoengineering" focus on active measures to "undo" the effects of human activities.  In particular, they focus on counteracting the warming induced by higher greenhouse gas concentrations \citep[e.g.][]{LentonVaughan2009}.  What seems like a logical consequence of these free energy considerations is that we first need to acknowledge that human activity will result in unavoidable impacts \citep[see also e.g.][]{Allenby2000}, but that the resulting impacts should be "engineered" towards enhancing the ability of the planet to generate free energy.  Given the sheer magnitude of human activity in terms of free energy consumption and its expected increase in the future, it would seem that such a form of Earth system engineering is the only sustainable way to support such an increase in human activity without depleting the sources of free energy generation within the Earth system.  Such active and conscious intervention would, of course, require careful analysis.

\section{What is needed in Earth system models to adequately represent free energy generation, dissipation and transfer?}
\label{sec:models}

To evaluate the effect of the different options of how humans appropriate free energy from the Earth system, one would need Earth system models that capture the dynamics of free energy generation, transfer, and dissipation.  Such an analysis, however, is practically absent in current applications of Earth system models.  While the energetics of some Earth system processes are diagnosed, for instance the kinetic energy generation in the atmosphere \citep{Lorenz1955} or the ocean \citep{FerrariWunsch2009}, most analyses focus on the balance of heating and cooling terms and associated temperature differences \citep[e.g. for the case of global warming, see][] {IPCC2001}.  For differences in heat content and temperature, this is justified, because the magnitude of free energy generation, transfer, and dissipation is much smaller than mean heating terms, so that their effect in the energy balance can be neglected.  However, it is the dynamics of free energy generation, transfer, and dissipation that shapes disequilibrium, and it is this disequilibrium that seems an appropriate way to characterize a habitable planet.  So it would seem that we miss a critical aspect when diagnosing Earth system model simulations of global change.

To allow for the analysis of the free energy balance, it would seem that some aspects are being missed in current Earth system models regarding the diagnostics, and, more importantly, regarding the adequate representation of free energy dynamics.  In terms of process representation, some processes are treated as diffusive even though they are not.  In contrast to a diffusive process, a non-diffusive process is associated with free energy generation, dissipation and, potentially, transfer to another process.  This is, for instance, the case for preferential flow of water in soils.  Preferential flow describes the rapid flow of water along spatially connected flow paths of minimum flow resistance within the soil, and this results in the faster depletion of gradients than would be predicted by using common, diffusion-based models \citep{ZeheEtal2010}.  Missing the dynamics of kinetic energy generation associated with the rapid flow does not allow for work possibly being drawn from the flow to transport material that could generate and maintain these connected flow paths within the soil.  Overall, without the dynamics of free energy generation, transfer, and dissipation, the dynamics of water movement would likely be slower, less dissipative, and hence misrepresented.

Thermodynamics and thermodynamic maximization can potentially also help to address the representation of heterogeneity and its effects on the functioning at larger scales in climate models.  Heterogeneity inherently reflects gradients, and as such reflects thermodynamic disequilibrium.  One example where thermodynamic maximization has been used to parameterize the effect of subgrid-scale heterogeneity is the case of pattern formation of vegetation in semiarid regions.  \citet{SchymanskiEtal2010} used a simple model based on a reaction-diffusion equation to show that the formation of patterns results in a thermodynamic maximization associated with water redistribution and biotic productivity.  It has also been hypothesized that the patterns found in sand dunes reflect maximum aeolean transport \citep{RubinHunter1987}, that is, the maximum transfer of kinetic energy of boundary layer winds to perform work on sand transport.  These studies provide a clue of the potential use of thermodynamic maximization to characterize the non-randomness of sub-grid scale heterogeneity and how it could possibly be described at a larger scale.  This would need further evaluations.

Another aspect that is likely absent in some model implementations is the adequate representation of free energy transfer.  When, for instance, air flows over an open water surface, some of the momentum gradient between the airflow and the water surface is transferred to generate water flow.  This momentum is not dissipated by turbulence, but rather transferred into another form of free energy in form of kinetic energy of the moving water.  Hence, the acceleration of the water surface is another form of reducing the momentum gradient between the airflow and the water surface.  The same example would also apply to the surface-atmosphere exchange on land, where momentum from the air flow is transferred to lift dust or to sway canopies.  When this form of momentum reduction is not accounted for and momentum gradients are only reduced by turbulent dissipation, this should result in an overestimation of the associated turbulent fluxes that exchange heat and matter across the surface.

Apart from ensuring thermodynamic consistency, adequately representing the dynamics of free energy generation, transfer and dissipation in Earth system models would allow us to:
\begin{itemize}
\item{test how close natural processes operate near maximum efficiency of free energy generation and transfer.  This could potentially result in simpler and better founded model parameterizations;}
\item{phrase interactions and feedbacks within the Earth system in the context of free energy generation and transfer.  This would then help us to better understand the processes and dynamics that are involved in maintaining maximum efficiency;}
\item{quantify the role of the biota in the generation of free energy by the Earth system.  This would allow for a better quantification of how much biotic activity contributes to the planetary disequilibrium of the Earth;}
\item{quantify the impact of human activity on the generation of free energy by Earth system processes.  This would potentially yield a better way to measure the impact of human activity than changes in surface temperature;}
\item{better quantify the free energy budget.  Because this budget includes the generation rates of free energy, it would provide a baseline budget for the availability of different forms of renewable energy;}
\item{determine strategies for future free energy appropriation by humans with minimum impact on free energy generation by Earth system processes.  This would ensure that the impacts of human activity on the Earth system has no detrimental effects on free energy generation by Earth system processes.}
\end{itemize}
It would thus seem that an adequate representation of free energy dynamics in Earth system models is not just a matter of evaluating the theory proposed here.  It would rather seem that free energy dynamics, especially regarding the chains of conversions of free energy from one form to another -- as in the example of momentum transfer at the surface above --, are a fundamental aspect that needs to be properly accounted for in models that aim to represent the non-equilibrium dynamics of the Earth system and its response to change.

\section{Summary and Conclusions}
\label{sec:summary}

I provided a holistic description of the functioning of the whole Earth system that is grounded in the generation, transfer and dissipation of free energy from external forcings to geochemical cycling and the associated fundamental limits to these rates.  Since free energy generation is needed to maintain a disequilibrium state, this description allows us to understand why the Earth system is maintained far from equilibrium without violating the second law of thermodynamics.  I showed how biotic activity generates substantial amounts of chemical free energy by exploiting free energy in solar photons that is not accessible to purely physical heat engines.  Hence, Lovelock's notion of chemical disequilibrium within the Earth's atmosphere as a sign for widespread life can be substantiated and quantified.  This paper can hence be seen as a direct continuation of the work by James Lovelock (\citeyear{Lovelock1965,Lovelock1975}) on understanding the Earth as a single system that is strongly shaped by life.

The relevance of this holistic description of the Earth system becomes apparent when the impacts of human activity is evaluated from this perspective. When using free energy consumption as a measure for human activity, it is evident that human activity as an Earth system process is far greater and significant in comparison to natural processes than what it would seem using other, more traditional measures.  An increase in human activity in the future, e.g. in terms of population size and standard of living, would inevitably result in greater needs for free energy generation.  If these needs are met by appropriating renewable sources of free energy within the Earth system then this is inevitably going to leave its impact in that the processes associated with this form of free energy will become less intense and "slow down".  The only sustainable way to meet the increasing needs for free energy by human activity would seem to use human technology in such a way that it would enhance the overall ability of the Earth system to generate free energy.  This was illustrated using the two examples of "desert greening" and the direct use of solar energy by photovoltaics or by heat engines using direct solar radiation in deserts.  Even though this would require careful analysis and planning of potential, detrimental side effects, it would seem that it is only through the large-scale use of human technology that the Earth system could sustainably generate more free energy, yielding a more prosperous and empowered future of the planet.

The evaluation of such potential scenarios of the planetary future associated with growing human activity with respect to their impact on the ability of the Earth system to generate free energy is at this point qualitative and would need to be more rigorously evaluated.  This can in principle be done by using climate models.  However, while these models are routinely used to predict typical meteorological properties such as temperature and precipitation rates, the evaluation in terms of rates of free energy generation and transfer are rare.  Apart from providing fundamental insights, thermodynamic maximization could provide a useful approach to capture sub-grid scale heterogeneity in global scale models.  The transfer of free energy among different variables would need to be adequately represented in these models to allow for the thermodynamic consistency of the dynamics.  Such thermodynamic consistency would, of course, be a first requirement for an improved prediction of human impacts and the planetary future.

\section{Acknowledgements}
The author acknowledges funding from the Helmholtz Alliance ÒPlanetary Evolution and LifeÓ.  This manuscript resulted from a presentation and subsequent discussions at the Newton Institute in August 2010.  The author thanks the organizers for the stimulating meeting.  Fruitful discussions with members of the Biospheric Theory and Modelling group are also acknowledged.  The author thanks two reviewers for their constructive and stimulating comments.


\bibliographystyle{dcu}


\newpage

\end{document}